\documentstyle[11pt,epsf]{article}

\begin{document}

\twocolumn[\hsize\textwidth\columnwidth\hsize\csname @twocolumnfalse\endcsname

\centerline{\bf Microcanonical Transfer Matrix Study 
of the $Q$-state Potts Model}
\vskip 3mm
\centerline{Richard J. Creswick and Seung-Yeon Kim}
\vskip 1mm
\centerline{\small \it Department of Physics and Astronomy,
University of South Carolina, Columbia, South Carolina 29208, USA}

\begin{abstract}
The microcanonical transfer matrix is used to study 
the zeros of the partition 
function of the $Q$-state Potts model. 
Results are presented for the Yang-Lee zeros of the 3-state model,
the Fisher zeros of the 3-state model in an external field $H_q<0$,
and the spontaneous magnetization of the 2-state model.
In addition, we are able to calculate the ground-state entropy 
of the 3-state model and find $s_0=0.43153(3)$ in excellent agreement
with the exact value, 0.43152...  
\end{abstract}
\vskip2mm]


{\noindent \bf I. Introduction}
\vskip 1mm

The $Q$-state Potts model\cite{potts} in two dimensions exhibits 
a rich variety of critical behavior and is very fertile
ground for the analytical and numerical investigation
of first- and second-order phase transitions.
With the exception of the $Q=2$ Potts (Ising) model
in the absence of an external magnetic field\cite{onsager},
exact solutions for arbitrary $Q$ are not known.
However, some exact results have been established for the ferromagnetic 
Potts model\cite{potts}.
For $Q\le4$ there is a second-order phase transition,
while for $Q>4$ the transition is first order.

The bond-energy for the $Q$-state Potts model is (in dimensionless units)
$$
E=\sum_{<i,j>}[1-\delta(\sigma_i,\sigma_j)],\eqno{(1)}
$$
where $<i,j>$ indicates a sum over nearest-neighbor pairs,
$\sigma_i=0$,...,$Q-1$,
and $E$ is a positive integer $0\le E\le N_b$,
where $N_b$ is the number of bonds on the lattice.
In addition to the energy there are $Q$ order parameters
$$
M_q=\sum_{k}\delta(\sigma_k,q),\eqno{(2)}
$$
where $q$ is a fixed integer between 0 and $Q-1$.
Note that $0\le M_q\le N_s$ is also an integer and 
$N_s$ is the number of sites on the lattice. 

If we denote the density of states with energy $E$ by $\Omega(E)$,
then the partition function for the Potts model is
$$
Z(y)=\sum_{E}\Omega_Q(E)y^E,\eqno{(3)}
$$
where $y=e^{-\beta}$. From (3) it is clear that $Z(y)$ is simply 
a polynomial in $y$.
If we wish to study the partition function in an external field
which couples to the order paramter, (2), then one needs to
enumerate the states with fixed energy $E$ and
fixed order parameter $M$, $\Omega(M,E)$. The partition function
in a magnetic field $H_q$ is again a polynomial given by
$$
Z(x,y)=\sum_{M=0}^{N_s}\sum_{E=0}^{N_b}\Omega_Q(M,E)x^M y^E,\eqno{(4)}
$$
where $x=e^{\beta H_q}$.


\vskip1mm
{\noindent \bf II. Microcanonical transfer matrix}
\vskip1mm

By the microcanonical transfer matrix ($\mu$TM)\cite{stoic} it is possible
to obtain {\it exact} integer values for $\Omega(E)$ and $\Omega(M,E)$.
Here we describe briefly the $\mu$TM on an $L\times L$ lattice with 
cylindrical boundary conditions.
First, an array, $\omega$, which is indexed by the energy $E$ 
and $L$ variables $\sigma_i$, $0\le i\le L-1$ for the first row
of sites is initialized as
$$
\omega(E,\sigma_0,\sigma_1,...,\sigma_{L-1})=
\delta(E-\sum_{i=0}^{L-1}h(\sigma_i,\sigma_{i+1})),\eqno{(5)}
$$
where $h(q,p)$ is the contribution to the energy from a single bond.

Following the Binder algorithm\cite{binder}, each spin in the row
is traced over in turn, introducing a new spin variable from
the next row,
$$
\omega(E;\sigma'_0,...,\sigma_{L-1})=
\sum_{\sigma_0}\omega(E-h(\sigma_0,\sigma'_0);\sigma_0,...,\sigma_{L-1}).
\eqno{(6)}
$$
This procedure is repeated until all the spins in the first row
have been traced over, leaving a new function of the $L$ spins
in the second row. The horizontal bonds connecting the spins 
in the second row are then taken into account by shifting
the energy,
$$
\omega(E;\sigma'_0,...,\sigma'_{L-1})=
\omega(E-\sum_{i=0}^{L-1}h(\sigma'_i,\sigma'_{i+1});
\sigma'_0,...,\sigma'_{L-1}).\eqno{(7)}
$$
This procedure is then applied to each row in turn until the final
row is reached. The density of states is then given by
$$
\Omega(E)=\sum_{\sigma'_0}\sum_{\sigma'_1}...\sum_{\sigma'_{L-1}}
\omega(E;\sigma'_0,\sigma'_1,...,\sigma'_{L-1}).\eqno{(8)}
$$

\begin{figure}
\epsfxsize=9cm
\epsfbox{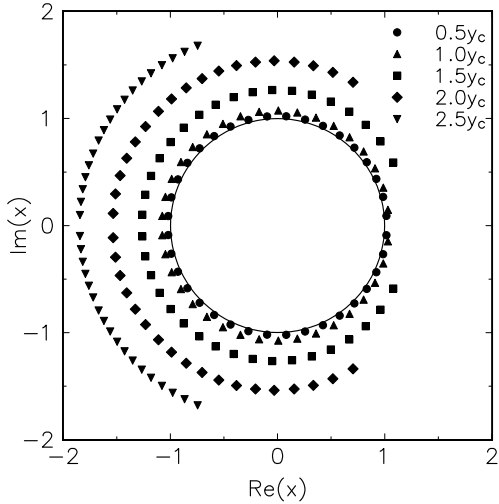}
\caption{Zeros of a $6\times6$ three-state Potts model
in the complex $x$ plane for several values of $y$ 
(cylindrical boundary conditions).}
\end{figure} 


\vskip1mm
{\noindent \bf III. Partition function zeros}
\vskip1mm

The analytic structure of the partition function is completely
determined by its Yang-Lee (YL) zeros in the complex $x$ plane\cite{yang}
and its Fisher zeros in the complex $y$ plane\cite{fisher}.
Using the $\mu$TM we have studied the YL zeros of the two-dimensional
Potts model for $Q=2$, for which the circle theorem of Lee and Yang\cite{yang}
applies, and, for the first time, for $Q>2$\cite{kim1}.
Figure 1 shows the YL zeros for the three-state Potts model.
At $y=0.5y_c$ the zeros are uniformly distributed close to the 
unit circle. As the temperature is increased the edge singularity
moves away from the real axis and the zeros detach from the 
unit circle. Finally, as $y$ approaches unity, the zeros converge 
on the point $x=1-Q$\cite{kim1}.
The $Q$-state model in an external field $H_q<0$ is in the same
symmetry class as the $Q-1$ state model.
We have studied the Fisher zeros of the 3-state model as a function
of $H_q$\cite{kim2} and determined the critical point and exponents.
The critical line is shown in Figure 2.

By introducing the density of zeros $g(\theta,y)$ Lee and Yang\cite{yang}
showed that the spontaneous magnetization of the Ising
model is given by
$$
m_0(y)=2\pi g(0,y).\eqno{(9)}
$$
Even though they have calculated the density of zeros of the
one-dimensional Ising model and confirmed the well-known fact that
the model does not exhibit ferromagnetism, the exact form of the density
of zeros is not known for $d\ge2$.
We have studied the density of zeros of the two-dimensional Ising
model numerically\cite{creswick}, and using (9) we have calculated 
the spontaneous magnetization (Figure 3) which agrees very well 
with the exact result\cite{onsager}.

\begin{figure}
\epsfxsize=9cm
\epsfbox{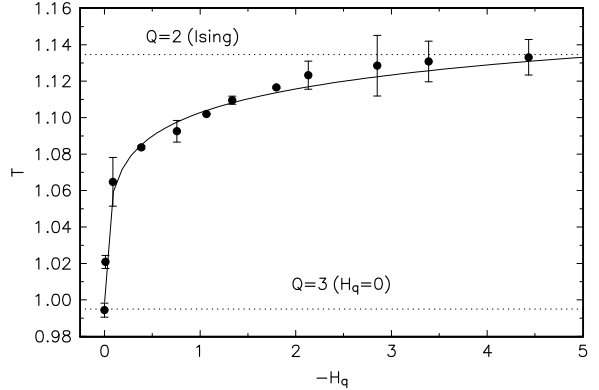}
\caption{Critical temperatures of the three-state Potts ferromagnet
as a function of the magnetic field. The upper dotted line is the Ising
transition temperature in the limit $H_q\to-\infty$, while the lower
dotted line shows the critical temperature of the three-state Potts
model for $H_q=0$.}
\end{figure}   


\vskip1mm
{\noindent \bf IV. Ground-state entropy}
\vskip1mm

The antiferromagnetic Potts model (APM) is much less well understood
than its ferromagnetic counterpart. One of the most interesting 
properties\cite{shrock} of this model is that for $Q>2$ the ground-state is
highly degenerate and the ground-state entropy is nonzero.
However, the exact ground-state entropy of the APM is not known
except for $Q=3$, where the ground-state entropy per site is\cite{lieb}
$$
s_0={3\over2}{\rm ln}{4\over3}=0.43152...\eqno{(10)}
$$

\begin{figure}
\epsfxsize=10cm
\epsfbox{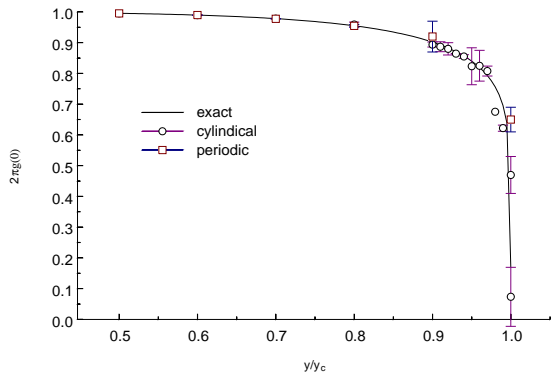}
\caption{Extrapolated density of zeros as a function of $y$.
The solid curve is Yang's exact result for the spontaneous magnetization. 
Open circles are our results for 
cylindrical boundary conditions, and open squares 
are for periodic boundary conditions.}
\end{figure}

From the density of states we have obtained the exact ground-state
entropy for finite lattices. Table 1 shows the ground-state degeneracy
$\Omega_0(L)$ of the $L\times L$ three-state APM.
The ground-state entropy is then given by
$$
s_0(L)={{\rm ln}\Omega_0(L)\over N_s}.\eqno{(11)}
$$
By using the Bulirsch-Stoer (BST) algorithm\cite{bst},
we have extrapolated our results for finite lattices to infinite size.
The BST estimate for $Q=3$ is $s_0=0.43153(3)$, in excellent
agreement with (10).


\vskip1mm
{\noindent \bf V. Discussion}
\vskip1mm

Using the $\mu$TM, we have calculated the restricted density of states
$\Omega(M,E)$ from which the partition function at any temperature
and in an arbitrary magnetic field can be evaluated.
We have studied the Yang-Lee zeros of the 3-state model and demonstrated
that, unlike the Ising model, the locus of zeros is not a circle.
Following Lee and Yang we have used the density of zeros to calculate 
the spontaneous magnetization for the 2-state model
and we are currently extending this work to $Q>2$.
Since the zeros of the partition function encompass both the
ferromagnetic and antiferromagnetic regimes, we are currently
investigating the critical behavior of the antiferromagnetic Potts model.
Also, the $\mu$TM is easily extended to three-dimensional lattices,
and we are now extending all of the above calculations to cubic lattices. 

Recently it has been suggested\cite{huller} that the microcanonical 
entropy may exhibit weaker finite-size effects
than the free energy.
The microcanonical ensemble also facilitates the study of phase coexistence
and the study of phase transformations in cluster and nuclear 
physics\cite{huller}.
The $\mu$TM method yields exact values for the microcanonical entropy
of finite systems allowing us to study finite size effects in the
microcanonical ensemble.

\begin{table}
\caption{The ground-state degeneracy of the three-state APM 
for even-size lattices (cylindrical boundary conditions).}
\vskip1mm
\begin{tabular}{rr}
\hline
$L$ &$\Omega_0(L)$ \\
\hline
4                                     &4626   \\
6                                 &37284186   \\
8                            &9527634436194   \\
10                    &77048019386429374638   \\
12           &19698820973096872077077373450   \\
14 &159147870862104841838351532192943853490   \\
\hline
\end{tabular}
\end{table}


\end{document}